\documentclass[revsymb,aps,prb,10pt]{revtex4}

\begin{document}
\title{Spiral vortices in 2D ferromagnet.}
\author{A.B. Borisov}
\address{Institute of Metal Physics, 620219, Ekaterinburg, Russia}
\author{I.G.Bostrem, A.S. Ovchinnikov}
\address{Department of Physics, Ural State University, 620083, Ekaterinburg, Russia}
\date{\today }

\begin{abstract}
We present a study of a new class of exact solutions having a form of spiral
vortices for an isotropic two-dimensional Heisenberg ferromagnet using a
continuum theory and direct numerical simulations of the spin system on a
square lattice. We find their features issued from the conservation laws and
describe their interaction. Reasons behind the formation of the proper spin
configurations on a square lattice are investigated.
\end{abstract}

\pacs{PACS numbers: 75.10.Hk, 74.78.Na, 75.40.Mg}
\maketitle

\section{ Introduction.}

In the last two decades solitons, vortices and other nonlinear excitations
in low-dimensional magnets have attracted a great interest of researchers.
These excitations play an essential role in two-dimensional (2D) magnetism
and contribute to breaking of the long-range order in 2D magnets. Magnetic
vortices are important for the dynamical and thermodynamical properties of
magnets, for a review see Refs. \cite
{Nonlinearity1,Nonlinearity2,Fluctuation}. Predictions of the
Belavin-Polyakov theory \cite{Belavin} for localized structures with a
finite energy (instantons) observed much later experimentally \cite{Barrett}
gave rise to intensive investigations of solitons in 2D magnets \cite
{Takeno,Bishop,Kolezhuk,Kosevich}.

Some years ago magnetic vortices have been directly observed in permalloy 
\cite{perm1,perm2,perm3,perm4,perm5} and Co magnetic nanodots \cite
{Cob1,Cob2,Cob3}. High frequency dynamical properties of the vortex state
magnetic dots have been probed by Brillouin light scattering of spin waves 
\cite{Light} and X-ray imaging technique \cite{xray}. In recent experiments
the spin-wave modes excited by magnetic field pulses of small
litographically define disks with a spin vortex configuration are imaged
using time-resolved magneto-optic Kerr microscopy \cite{perm6} and phase
sensitive Fourier transformation technique \cite{Fourier}. Both axially
symmetric dynamical modes showing concentric nodes and symmetry breaking
azimuthal eigenmodes having azimuthal nodes have been observed. An analysis
of the time and frequency dependencies \cite{PRL2005} of the modes
demonstrates that for moderate field pulses and large magnetic elements
(several tens of microns) the excitation spectrum is dominated by
magnetostatic modes. However, as noted by authors \cite{PRB2005}, when the
size of the elements is reduced or higher modes are excited, the exchange
interactions can, in general, no longer be ignored and the dynamic response
gradually changes from a purely magnetostatic to an exchange-dominated one.
One of the aims of the paper is to show an existence of nonlinear modes both
with circular and azimuthal nodes in a 2D isotropic ferromagnet obtained
with an account only exchange interaction. These modes can be observed as
spiral-like vortex configurations. Besides a detailed study of the spiral
solutions within both XY and Heisenberg models  is of interest for the
possible applications in the physics of liquid crystals, quantum Hall effect
and in the study of biological systems featuring self-organized spiral
structures \cite{Gross,Nedelec}.

In this paper, we present a study of the spiral vortices in the isotropic 2D
ferromagnet using a continuum theory and direct numerical simulations of the
spin system on a lattice. Our principal concern is to understand in detail
the structure of the spiral patterns and to find out reasons of their
appearance. Real compounds are not ideal systems and lattice defects such as
impurities, local fields and anisotropy are present in any material sample.
The effect of nonmagnetic impurities (vacancies) on vortex and
vortex-antivortex structures has recently been studied for 2D magnetic
models with XY symmetry \cite{Pires,Leonel,Wysin}. These investigations show
that an ideal vortex or their pair formations are deformed if the vortex
centers are near the vacancy. In the continuum limit an account of the
imputities results in logarithmic singularities in the spin field. On the
other hand, in the quantum field studies of 2D $O(N)$ models enjoying a
continuous symmetry these classical configurations with the logarithmic
singularities are known as ''superinstantons'' \cite{Seiler}. To produce
these configurations a novel ''superinstanton'' boundary conditions (SIBC)
were introduced. These consist of Dirichlet conditions on the boundary of
the system, and the additional freezing of one spin in the center of the
sample. It was argued that unlike to standard free, periodic, and Dirichlet
boundary conditions SIBC do not possess a well defined perturbation
expansion \cite{Niedermayer,Aguado} that means that by fixing the spin in
the center one change a ground state of the system.

The paper is organized in the following way. In Sec. II the continuum
approximation based upon equations of nonlinear spin dynamics is presented.
We briefly review existing literature and show that the continuum isotropic
Heisenberg model yields two types of static solutions. A new class of exact
solutions of the model, that are local minima of the classical energy, is
obtained using a special linearization procedure. We find a harmonic
function of initial dynamical variables obeying the linear Laplace equation.
Then, an inverse transformation gives solutions of the initial nonlinear
model as functions of the harmonic solutions of the Laplace equation. As a
result, new types of exact solutions for 2D isotropic ferromagnet are
generated where spiral vortices are of special interest. They are likely to
be relevant for non-perfect systems with defects. Thus, it has been recently
shown that vortices are attracted by a nonmagnetic impurity \cite{Attract}.

In Sec. III we consider numerical simulations of spin configurations
predicted by the continuum theory. Firstly we investigate spin textures for
the planar $xy$-model with the imposed SIBC. We show that in the sector with
topological charge $q=0$ the ground state is a logarithmic source of the
strength $\alpha $. We find how the strength depends on a turn of the
\thinspace fixed spin in regard to spins at the system edges and check the
continuum theory prediction for the energy. We consider the simplest
formation of such kind of logarithmic sources, a pair including two sources
of opposite strengths $\alpha $ and $-\alpha $. Then we investigate the case
when the structure takes an out-of-plane form. A numerical simulation gives
the structures which are close to the nodal solutions of Heisenberg model.
In the topological sectors $q\neq 0$ the ground state is either a planar
logarithmic spiral ($xy$-model) or a space spiral vortex with an
out-of-plane form (Heisenberg model). We show that these spin configurations
minimizing an energy with imposed SIBC well reproduce features predicted by
the continuum theory.

\section{ Analytical results.}

The model to be investigated is the isotropic spin-$S$ Heisenberg
ferromagnet defined by the Hamiltonian

\begin{equation}
H=-\sum\limits_{p,n}J_{pn}\vec{S}_p\vec{S}_n  \label{Hamilka}
\end{equation}
where $\vec{S}_p$ represents the spin operator at the site $p$ of a 2D
square lattice with the nearest-neighbor distance $\vec{a}$ , and $%
J_{pn}=J\delta _{n,p+\vec{a}}$ ($J>0$) are the nearest neighbor exchange
couplings. The non-linear differential equations describing the dynamics of
the model can be obtained by taking diagonal matrix elements of the equation
of motion for the raising operator $S_p^{+}=S_p^x+iS_p^y$ 
\begin{equation}
-i\hbar \frac{dS_p^{+}}{dt}=\left[ H,S_p^{+}\right]  \label{motion}
\end{equation}
of the $p$-th spin in spin-coherent representation $\left| \Omega
\right\rangle =\prod\limits_p\left| \theta _p,\phi _p\right\rangle ,$ where $%
0\leq \theta _p\leq \pi $ and $0\leq \phi _p<2\pi $ parametrize the spin
states on the unit sphere \cite{Bala}. For the bilinear Hamiltonian this
results in the system for the classical variables $\{\theta _p,\varphi _p\}$
parametrizing the $\vec{S}_p$ spin 
\begin{equation}
\sin \theta _p\frac{\partial \varphi _p}{\partial t}=-\frac S\hbar
\sum\limits_nJ_{np}\sin \theta _n\cos \theta _p\cos (\varphi _p-\varphi
_n)+\sin \theta _p\frac S\hbar \sum\limits_nJ_{np}\cos \theta _n,
\label{System1}
\end{equation}

\begin{equation}
\frac{\partial \theta _p}{\partial t}=\frac S\hbar \sum\limits_nJ_{np}\sin
\theta _n\sin (\varphi _n-\varphi _p),  \label{System2}
\end{equation}
hereinafter, $n$ runs over the nearest neighbors. In the continuum limit we
introduce the fields $\theta (x,y)$, $\varphi (x,y)$, which are defined in
the $(x,y)$-plane. The equation of motion for static solutions $\frac{%
\partial \theta _p}{\partial t}=\frac{\partial \varphi _p}{\partial t}=0$
can be obtained by applying the continuum approximation to the equation of
spin motion on the discrete lattice. This yields 
\begin{equation}
\left\{ 
\begin{array}{c}
\triangle \theta =\sin \theta \cos \theta \left( \vec{\nabla}\varphi \right)
^2 \\ 
\vec{\nabla}\left( \sin ^2\theta \,\vec{\nabla}\varphi \right) =0.
\end{array}
\right.  \label{system}
\end{equation}

A remarkable property of these equations is a conformal invariance that
allows us to subdivide their static solutions into two groups. For the first
group of solutions the expression

\begin{equation}
\frac{\partial \theta }{\partial {x}}\frac{\partial \varphi }{\partial {y}}-%
\frac{\partial \theta }{\partial {y}}\frac{\partial \varphi }{\partial {x}}
\label{X2}
\end{equation}
does not equal to zero. Then we can obtain new types of solutions from
already known ones via conformal transformations. Indeed, let $\theta
_1(x,y),\varphi _1(x,y)$ are some particular solutions of Eqs. (\ref{system}%
). We may see by direct calculations that the fields $\theta
_1(u_1(x,y),u_2(x,y))$, $\varphi _1(u_1(x,y),u_2(x,y))$ are also the
solutions of the same system provided the $u_1+iu_2$ is an arbitrary
analytic function $F$ of the argument $x+iy$ 
\begin{equation}
u_1+iu_2=F(x+iy).  \label{X3}
\end{equation}

Until now, all known solutions belong to the first group. For this case $%
\varphi (x,y)$ may be written in the simple form 
\begin{equation}
\varphi (x,y)=u_2(x,y),  \label{X4}
\end{equation}
and the another function $\theta $ depends only on the $u_1(x,y)$, i.e. $%
\theta _1(u_1(x,y),u_2(x,y))=\theta _1(u_1)$, and obeys the simple equation
of pendulum motion

\begin{equation}
\theta _{u_1u_1}(u_1)=\frac 12\sin \left[ 2\theta (u_1)\right] .  \label{X5}
\end{equation}

The cases of the infinite and finite pendulum motions correspond to the
following solutions

\begin{eqnarray}
\cos \theta (x,y) &=&sn\left[ \frac{u_1(x,y)}k,k\right] \qquad (0<k<1),
\label{X7} \\
\cos \theta (x,y) &=&k\,sn\left[ u_1(x,y),k\right] \qquad (0<k<1),
\label{X8}
\end{eqnarray}
where the Jacobi elliptic function of modulus $k$ is used. Eqs.(\ref{X4},\ref
{X7},\ref{X8}) describe a set of quiescent topological defects centered at
positions $z_i=x_i+iy_i$ (see Ref. \cite{Borisov}) with 
\begin{equation}
u_1(x,y)+iu_2(x,y)=\sum_{i=1}^n\left( \frac{2i\,k\,K}\pi N_i+Q_i\right) \ln
(x+iy-z_i)\qquad (N_i,Q_i\in Z).  \label{X9}
\end{equation}
Here, $K=K(k)$ is the complete elliptic integral of the first kind, and $c_i$
are fixed complex parameters.

For $n=1$, $N_1=0$ and $k=1$ the solution of Eqs.(\ref{X4},\ref{X7},\ref{X8},%
\ref{X9}) 
\[
\cos \theta (x,y)=\tanh u_1(x,y),\;u_1=Q\ln \sqrt{x^2+y^2},\;u_2=Q\arctan
\left( \frac yx\right) 
\]
coincides with the Belavin-Polyakov vortex (''baby'' soliton) \cite{Belavin}
with the topological charge $Q$.

For $n=1$, $N_1\neq 0$, $k\neq 1$ the solution of Eqs.(\ref{X7},\ref{X9})
represents a $N_1$-armed logarithmic spiral consisting of $2N_1$ spiral
regions separated by the same number of logarithmic spiral walls \cite
{Borisov}. The field $\varphi (x,y)$ (\ref{X4},\ref{X9}) determined by the
topological charge $Q_1$ does not alter along the logarithmic spiral curves
in the $(x,y)$-plane. Note that the spiral vortices in the ferromagnet,
involving both $\theta $ and $\varphi $ variables, have the different
mathematical structure in comparison with the optical spiral vortices and
the spiral vortex solution of the complex Landau-Lifshitz model \cite{Pismen}
where only $\varphi $ angle is used to build proper configurations.

We are more interested in the second group of solutions when the expression (%
\ref{X2}) equals to zero. In this case the angle $\varphi (x,y)$ is an
arbitrary function of $\theta (x,y)$. Then we use the ansatz $\vec{\nabla}%
\varphi =f(\theta )\vec{\nabla}\theta $ to find solutions of this class with 
$\vec{\nabla}\varphi ||$ $\vec{\nabla}\theta $. After eliminating the $\vec{%
\nabla}\varphi $ from [Eq.(\ref{system})], the equation for $\theta $ being

\begin{equation}
\sin \theta \frac{df}{d\theta }+2f\cos \theta +f^3\sin ^2\theta \cos \theta
=0.  \label{Bernul}
\end{equation}
This is Bernoulli equation in the variable $f(\theta )$, the general
solutions are therefore given by

\begin{equation}
f(\theta )=\frac 1{\sqrt{{c}^2\sin ^4\theta -\sin ^2\theta }},
\label{fsolut}
\end{equation}
where $c^2>1$ is an arbitrary parameter. Inserting the ratio $\vec{\nabla}%
\varphi =f(\theta )\vec{\nabla}\theta $ and Eq.(\ref{fsolut}) into Eq.(\ref
{system}) we find the fields $\theta (\vec{r})$, $\varphi (\vec{r})\,$ as 
\begin{equation}
\left\{ 
\begin{array}{c}
\cos \theta =\frac{\sqrt{c^2-1}}c\cos a, \\ 
\varphi =\arctan \left( c\tan a\right) +\varphi _0.
\end{array}
\right. ,  \label{mainsol}
\end{equation}
where the field $a(x,y)$ is satisfied the Laplace equation 
\begin{equation}
\triangle a=0.  \label{X10}
\end{equation}

We will only consider special case of solution for $a(x,y)$ 
\begin{equation}
a=\sum_{i=1}^n\alpha _i\ln \left( \frac{\sqrt{(x-x_{0i})^2+(y-y_{0i})^2}}{R_i%
}\right) +\sum_{i=1}^nq_i\arctan \left( \frac{y-\tilde{y}_{0i}}{x-\tilde{x}%
_{0i}}\right) ,\quad q_i\in Z.  \label{manysol}
\end{equation}
with the parameters $\alpha _i$, $q_i$, $R_i$, $c$. In the above expression,
($x_{0i}$, $y_{0i}$) and ($\tilde{x}_{0i}$, $\tilde{y}_{0i}$) are positions
of sources and vortices, respectively, $\varphi _0$ is an initial value of
azimuthal angle $\varphi $. We call the parameter $\alpha $ a strength of
source that is identical to the term used in the hydrodynamic theory. From
Eq.(\ref{mainsol}) we see that the parameter $c$ governs out-of-plane spin
components. In the soliton (\ref{mainsol}) the spins are confined to the
vicinity of the $xy$-plane with $\pi /2-\theta _{max}\leq \theta <\pi
/2+\theta _{max}$ unlike the solutions of Eqs. (\ref{X4},\ref{X7},\ref{X9}).
The maximal value $\theta _{max}$ is given by $\theta _{max}=\arcsin \sqrt{%
c^2-1}/c$ for $n=1$. Eqs.(\ref{mainsol},\ref{manysol}) include both novel
and well-known solutions considered early by some authors. Below, we list
these cases.

1) $n=1$, $c=1$ (pure in-plane solutions). Using Eq.(\ref{mainsol}) we find
immediately that $\theta =\pi /2$ and $\varphi =q\phi +\alpha \ln (r/R)$
written in the polar coordinates $r,\phi $. For $\alpha =0$ we restore
Kosterlitz-Thouless (KT) vortices, and for $q=0$ we have ''sources'', the
equation for $\varphi $ being $\varphi =\alpha \ln (r/R)$ \cite
{Cvelik,Seiler}.

2) $n=1$, $\alpha \neq 0$, $q=0$ (solutions with out-of-plane spin
components). Eq.(\ref{mainsol}) can be written as

\[
\cos \theta =\frac{\sqrt{c^2-1}}c\cos \left( \alpha \ln \frac rR\right) , 
\]
\begin{equation}
\varphi =\arctan \left( c\tan \left( \alpha \ln \frac rR\right) \right)
+\varphi _0  \label{targets}
\end{equation}
This agrees with the result obtained in Ref. \cite{Bostrem1} (''nodal''
solutions).

A new class of exact solutions 
\[
\cos \theta =\frac{\sqrt{c^2-1}}c\cos \left( \alpha \ln \frac rR+q\phi
\right) , 
\]
\begin{equation}
\varphi =\arctan \left( c\tan \left( \alpha \ln \frac rR+q\phi \right)
\right) +\varphi _0  \label{spirus}
\end{equation}
are the two-dimensional spirals \cite{Bostrem}. Figure 1 presents them for
different parameters $\alpha $ and $q$. The former value assigns a spiral
twist and the latter defines a number of spiral arms. For $\alpha =0$ we
obtain a vortex with a non-zero out-of plane component

\[
\cos \theta =\frac{\sqrt{c^2-1}}c\cos \left( q\phi \right) , 
\]
\begin{equation}
\varphi =\arctan \left( c\tan \left( q\phi \right) \right) +\varphi _0.
\label{outvort}
\end{equation}
Unlike the Skyrmion the soliton has a zero topological charge $\pi _2(S_2)=0$%
. Since $\theta $ does not depend on the radial coordinate $r$, the solution
has no axial symmetry.

There are several conserved quantities that are important in what follows:
the total energy $E$, the linear momentum $\vec{P}$, the angular momentum $%
L_z$ and the total number of spin reversals $N$ (see, e.g., Ref. \cite
{Egorov}).

The energy of the soliton given by Eqs.(\ref{mainsol}) can be evaluated in
continuum approximation resulting in a more compact form

\begin{equation}
E=\frac 12JS^2\int d\vec{r}\left( \vec{\nabla}a\right) ^2.  \label{energy}
\end{equation}
Using Eq.(\ref{manysol}) for the function $a$ we obtain

\begin{equation}
E=\pi JS^2\left\{ \sum_i\left( \alpha _i^2+q_i^2\right) \ln \frac L{r_0}%
+\sum_{ij}\left( \alpha _i\alpha _j+q_iq_j\right) \ln \frac L{d_{ij}}%
\right\} ,  \label{enervort}
\end{equation}
$d_{ij}=\sqrt{(x_{0i}-x_{0j})^2+(y_{0i}-y_{0j})^2}$ is a distance between $i$%
-th and $j$-th vortices, $L$ is a size of the system and $r_0$ is a cut-off
radius where the continuum approximation breaks down. After the formal
substitution $q_iq_j\rightarrow \alpha _i\alpha _j+q_iq_j$ in Eq.(\ref
{enervort}) we recognize the energy of interacting in-plane vortices. We
note the absence of terms $\alpha _iq_j$. One can see that vortices and
sources do not interact with each other. The parameters $R_i$ and $c$ do not
enter into the expression at all.

Two spiral vortices of opposite values ($\alpha $, $q$) and ($-\alpha $, $-q$%
) with a pair separation $d$ has the finite energy 
\[
E=2\pi JS^2\left( \alpha ^2+q^2\right) \ln \left( \frac d{r_0}\right) 
\]
meaning that these vortices may be bound in pairs.

Next we derive another conserved quantities related with the spiral vortex.
The density of momentum is determined by the formula

\[
\vec{P}=\frac{\hbar S}{c+\sqrt{c^2-1}\cos a}\vec{\nabla}a 
\]
consisting of the radial

\begin{equation}
P_r=\frac{\hbar S}{c+\sqrt{c^2-1}\cos a}\frac \alpha r,  \label{Pr}
\end{equation}
and the azimuthal

\begin{equation}
P_\phi =\frac{\hbar S}{c+\sqrt{c^2-1}\cos a}\frac qr  \label{Phi}
\end{equation}
parts. The $P_r$ and $P_\phi $ components are defined by the twist parameter 
$\alpha $ and the vorticity $q$, respectively. Substituting (\ref{Pr}, \ref
{Phi}) in the relations

\[
\int\limits_0^{2\pi }P_r R_{0}d\phi =\alpha \hbar S\int\limits_0^{2\pi
}d\phi \frac 1{c+\sqrt{c^2-1}\cos \left( q\phi +\alpha \ln \frac rR\right) }%
=2\pi \alpha \hbar S 
\]
and

\[
\int\limits_0^{2\pi }P_\phi R_{0} d\phi =q\hbar S\int\limits_0^{2\pi }d\phi 
\frac 1{c+\sqrt{c^2-1}\cos \left( q\phi +\alpha \ln \frac rR\right) }=2\pi
q\hbar S 
\]
we clarify the physical meaning of the quantities $\alpha $ and $q$. The
first constraint is related with a flow of the momentum through the circle
of the radius $R_0$ surrounding the vortex core and the last one determines
a quantized circulation along the circle.

The total orbital angular momentum $L_z$ along the rotation axis through the
area $\pi L^2$ here reads as

\[
L_z=S\hbar q\int_0^Lrdr\int_0^{2\pi }d\phi \frac 1{c+\sqrt{c^2-1}\cos \left(
q\phi +\alpha \ln \frac rR\right) }=S\hbar q\,\pi L^2, 
\]
where the density of the angular momentum is defined as

\[
\frac{S\hbar q}{c+\sqrt{c^2-1}\cos \left( q\phi +\alpha \ln \frac rR\right) }%
. 
\]
The total linear momentum amounts here to $\int \vec{P}d^2r=0$.

The conservation of total number of spin reversals (magnon density)

\begin{equation}
N=S\left( 1-\frac{\sqrt{c^2-1}}c\cos \left( \alpha \ln \frac rR+q\phi
\right) \right)  \label{magden}
\end{equation}
involves the magnon density current

\begin{equation}
\vec{j}=\frac{JS^2}{\hbar c}\left( \frac qr\vec{e}_\phi +\frac \alpha r\vec{e%
}_r\right) .  \label{magcur}
\end{equation}
From Eqs.(\ref{Pr}, \ref{magcur}), it follows that the presence of
additional terms with a non-zero strength of source $\alpha $ produces
radial components in the densities of the momentum $\vec{P}$ and the magnon
current $\vec{j}$.

To complete our analytical study we discuss stability of the spiral
vortices. Given the energy (\ref{energy}) we would like to consider the
effect of perturbation $\delta \psi $, that belongs to the same class as the 
$a(x,y)$ does, on the soliton structure [Eq.(\ref{mainsol})]. This yields

\[
E[a+\delta \psi ]=\frac 12JS^2\int d\vec{r}\left( \vec{\nabla}a+\vec{\nabla}%
\delta \psi \right) ^2 
\]
\[
=\frac 12JS^2\int d\vec{r}\left( (\vec{\nabla}a)^2+2(\vec{\nabla}a)(\vec{%
\nabla}\delta \psi )+(\vec{\nabla}\delta \psi )^2\right) = 
\]
\[
=\frac 12JS^2\int d\vec{r}\left( (\vec{\nabla}a)^2+(\vec{\nabla}\delta \psi
)^2\right) >E[a], 
\]
i.e. the system pays an energy cost and the soliton turns out to be stable
against the small perturbations of the field $a(x,y)$.

Among the several questions that should be arisen in the above analysis, an
origin of logarithmic sources and a range of possible values of $\alpha $
parameter are ones of the most important. Due to the intrinsic interest, an
analysis of physical reasons behind the formation of spiral vortices is
called for. In addition, the continuum theory cannot completely describe
subtle differences occurring on the lattice at short length scales.
Therefore we should concern how to organize a numerical process leading to
spin configurations that may be compared with the spiral vortices. We are
aware of the difference between these spin patterns and spiral vortices
predicted by the continuum theory. In the first case we deal with a ground
state of the system under certain constrains and in the other case with
stationary nonlinear excitations. Our studies will involve lattice model on
2D square lattice. Ultimately, a lattice model is the original source of any
continuum theoretical description. We will then compare the continuum theory
predictions to results found in numerical calculations.

\section{NUMERICAL SIMULATIONS}

\subsection{The model.}

To describe in full detail the method of numerical simulations we rewrite
the system (\ref{System1},\ref{System2}) in the form convenient for an
iteration procedure. From Eq.(\ref{System2}) we get 
\[
\cos \varphi _p\left( \sum\limits_nJ_{np}\sin \theta _n\sin \varphi
_n\right) =\sin \varphi _p\left( \sum\limits_nJ_{np}\sin \theta _n\cos
\varphi _n\right) 
\]
that yields 
\begin{equation}
\sin \varphi _p=\pm \frac{\sum\limits_nJ_{np}\sin \theta _n\sin \varphi _n}{%
\sqrt{\left( \sum\limits_nJ_{np}\sin \theta _n\sin \varphi _n\right)
^2+\left( \sum\limits_nJ_{np}\sin \theta _n\cos \varphi _n\right) ^2}},
\label{sinpn}
\end{equation}

\begin{equation}
\cos \varphi _p=\pm \frac{\sum\limits_nJ_{np}\sin \theta _n\cos \varphi _n}{%
\sqrt{\left( \sum\limits_nJ_{np}\sin \theta _n\sin \varphi _n\right)
^2+\left( \sum\limits_nJ_{np}\sin \theta _n\cos \varphi _n\right) ^2}}
\label{cospn}
\end{equation}
and the upper sign must be taken for a ferromagnet ($J_{np}>0$). Similar
equation for $\theta _p$ is obtained from Eq.(\ref{System1}) which can be
written as 
\begin{equation}
\sin \theta _p\left( \sum\limits_nJ_{np}\cos \theta _n\right) =\cos \theta
_p\sum\limits_nJ_{np}\sin \theta _n\left( \cos \varphi _n\cos \varphi
_p+\sin \varphi _n\sin \varphi _p\right) .  \label{sincostp}
\end{equation}

Application of (\ref{sinpn},\ref{cospn}) to this equation gives 
\begin{equation}
\cos \theta _p=\sin \theta _p\frac{\sum\limits_nJ_{np}\cos \theta _n}{\sqrt{%
\left( \sum\limits_nJ_{np}\sin \theta _n\sin \varphi _n\right) ^2+\left(
\sum\limits_nJ_{np}\sin \theta _n\cos \varphi _n\right) ^2}}  \label{rat1}
\end{equation}
that after some simplifications yields the expression used in a numerical
algorithm 
\begin{equation}
\cos \theta _p=\frac{\sum\limits_nJ_{np}\cos \theta _n}{\sqrt{\left(
\sum\limits_nJ_{np}\cos \theta _n\right) ^2+\left( \sum\limits_nJ_{np}\sin
\theta _n\sin \varphi _n\right) ^2+\left( \sum\limits_nJ_{np}\sin \theta
_n\cos \varphi _n\right) ^2}}.  \label{costp}
\end{equation}
Together with (\ref{rat1}) it implies $\sin \theta _p>0$.

In actual practice, the spin configuration was found by using the original
lattice spin fields $\vec{S}_n$ and iteratively repointing each along the
effective local field due to its neighbors. Scanning linearly through the
lattice each site was updated in sequence, being reset along the net field
due partly to some unchanged neighbors and some that have already been
repointed. This gives fast convergence than a synchronized global update.
The iterations stop if the sum 
\begin{equation}
\sigma =\sqrt{\sum\limits_{i,j=0}^N\left( \sin \theta _{ij}^{(k)}-\sin
\theta _{ij}^{(k-1)}\right) ^2+\sum\limits_{i,j=0}^N\left( \sin \varphi
_{ij}^{(k)}-\sin \varphi _{ij}^{(k-1)}\right) ^2}  \label{toler}
\end{equation}
taken over a quarter of the lattice on the $k$-th step is less than
tolerance $10^{-6}\div 10^{-10}$. We employed the lattice coordinates in (%
\ref{toler}) for the notation of site indices.

The most difficult computational problem in carrying out this program is to
find the initial configuration that relaxes to a target spin configuration.
It is meaningful to impose appropriate boundary conditions too. Obviously
this a rich problem with a wide choice of options.

One way is to take the configuration according to continuum formula and
assume that each spin has small amplitude dynamic deviations from the
starting structure. Another approach has been used in a study of a single
magnetic vacancy centered in vortex \cite{Wysin}. For numerical calculations
a finite core circular system of radius $R_c$ is taken. Lattice sites are
set up surrounding the origin and only those within radius $R_c$ are kept. A
detailed discussion of starting configurations needed for a finding of
proper lattice structure and boundary conditions adopted in calculations
will be given in every case. We note here, they are essentially different
for vortex, logarithmic and spiral spin arrangements and their pairs.

Another important point of numerical simulations is a criteria of consistent
between the relaxed spin configuration and an appropriate continuum
solution. We suggest the following scheme for the comparing. (i) The
continuum theory is not relevant to spins at sites close to the vortex core.
Far from the core the relaxed spin angles must well be described by the
continuum formula. Thus, we have to control this coincidence with a
prescribed precision in a region where the continuum description works. (ii)
The number of independent parameters in a continuum solution must be the
same as a number of corresponding degrees of freedom controlled in numerical
simulations. (iii) A relaxed configuration should not lose a symmetry of
continuum solution, i.e. it should have a similar dependence on the space
coordinates ($r,\phi $). (iv) In addition, we confirm the finding of proper
solution by analyzing the total energy of a relaxed configuration

\[
E=\frac{S^2}2\sum\limits_{n,p}J_{np}\left[ \sin \theta _n\cos \theta _p\cos
(\varphi _p-\varphi _n)+\cos \theta _p\cos \theta _n\right] 
\]
comparing it with a continuum theory prediction.

\subsection{Logarithmic source}

\subsubsection{XY-model}

Our current aim is to perform simulations of a logarithmic source in the
planar $XY$ model. The static in-plane angles satisfy the discrete nonlinear
equations

\begin{equation}
\sin \varphi _p=\frac{\sum\limits_nJ_{np}\sin \varphi _n}{\sqrt{\left(
\sum\limits_nJ_{np}\sin \varphi _n\right) ^2+\left( \sum\limits_nJ_{np}\cos
\varphi _n\right) ^2}},  \label{xy1}
\end{equation}
\begin{equation}
\cos \varphi _p=\frac{\sum\limits_nJ_{np}\cos \varphi _n}{\sqrt{\left(
\sum\limits_nJ_{np}\sin \varphi _n\right) ^2+\left( \sum\limits_nJ_{np}\cos
\varphi _n\right) ^2}}  \label{xy2}
\end{equation}
drawn from Eqs.(\ref{sinpn},\ref{cospn}) after the substitution $\sin \theta
_n=1$.

A homogeneous arrangement $\varphi _p = \varphi = \mbox{const}$ is an
obvious solution of these equations for any set of the exchange couplings $%
J_{np}$. An attempt to carry out numerical simulations using a magnetic
vacancy with some zero nearest-neighbor exchange couplings, a magnetic
impurity with another spin and/or different exchange leads to the uniform
arrangement and does not permit to get a structure similar to a logarithmic
solution

\begin{equation}
\varphi =\varphi _0+\alpha \ln \frac rR.  \label{logasol}
\end{equation}
A close examination of an in-plane arrangement given by the analytical model
shows that there is some bending of spins in the center with reference to
spin order in the vicinity of the system edges. This nonuniformity of spin
distribution can gain insight into the numerical process driving the
starting spin configuration to desired in-plane structure. A reason behind
the formation of the nonuniformity may be either a local magnetic field or a
local anisotropy. We fix our attention on the first case. Indeed, an
inclusion into the Hamiltonian of the local field $h$, acting on a spin at
position $\vec{r}_0$, whose direction relative to fixed spins at boundary is
determined by the angle $\omega $ (see Fig.2) 
\begin{equation}
H_z=-g\mu _0hS\int d\vec{r}\,\delta (\vec{r}-\vec{r}_0)\cos \left[ \omega
-\varphi (\vec{r})\right]  \label{Zeem}
\end{equation}
leads to the continuum equation 
\begin{equation}
\triangle \varphi (\vec{r})=\frac{g\mu _0h}{JS}\sin \left[ \omega -\varphi (%
\vec{r})\right] \delta (\vec{r}-\vec{r}_0).  \label{nonunif}
\end{equation}
It then follows that 
\begin{equation}
\varphi (\vec{r})=\varphi _u(\vec{r})+\frac{g\mu _0h}{2\pi JS}\sin \left[
\omega -\varphi (\vec{r}_0)\right] \ln \left| \vec{r}-\vec{r}_0\right| ,
\label{nonunisol}
\end{equation}
where $\varphi _u$ is an arbitrary solution of the Laplace equation $%
\triangle \varphi _u=0$ in two dimensions. A comparison of the result with
Eq.(\ref{logasol}) yields 
\begin{equation}
\alpha =\frac{g\mu _0h}{2\pi JS}\sin \left[ \omega -\varphi (\vec{r}%
_0)\right] .  \label{alpha1}
\end{equation}
We can proceed and estimate the expression using consideration of the
mean-field theory. Taking $\omega =\pi /2$ and suggesting that all the spins
are aligned parallel to the one direction except the core spin, pointed
fixedly along an axis determined both the exchange field $zJS$ of the
nearest neighbors and the local filed $h$, we easily find 
\begin{equation}
\sin \left[ \omega -\varphi (\vec{r}_0)\right] =\frac{zJS}{\sqrt{\left(
zJS\right) ^2+\left( g\mu _0h\right) ^2}},  \label{sinal}
\end{equation}
where $z$ is the number of nearest neighbors. Therefore, 
\begin{equation}
\alpha =\frac z{2\pi }\frac{g\mu _0h}{\sqrt{\left( zJS\right) ^2+\left( g\mu
_0h\right) ^2}}.  \label{alphaMF}
\end{equation}
As we can see $\alpha $ ranges from $\ 0$ to $\frac z{2\pi }\approx 0.636$
when $h$ increases from zero to infinity. Most importantly, the local
magnetic field is the reason of an appearance of the logarithmic source in
the system and there is the upper limit for $\alpha $ values.

To check the predictions with the numerical simulation data we consider a
square lattice of size $(2N+1)\times (2N+1)$ shown in Fig.3 and take $%
J_{np}=J$ for simplicity. We carry out the iteration process beginning from
one of the corners of the lattice. Let the local magnetic field directed
along $j$-axis acts on a core spin which has the coordinates ($N,N$) [Fig.
4a]. This spin should be included into the numerical scheme [Eqs.(\ref{xy1}, 
\ref{xy2})] with a little modification 
\begin{equation}
\cos \varphi _{NN}=\frac{\sum\limits_n\cos \varphi _n+g\mu h/(JS)}{\sqrt{%
\left( \sum\limits_n\sin \varphi _n\right) ^2+\left( \sum\limits_n\cos
\varphi _n+g\mu h/(JS)\right) ^2}}.  \label{magfilit}
\end{equation}

During the iterations a turn of the central spin $\varphi (\vec{r}_0)$
proportional to the applied field $h$ is seen to develop in regard to
uniform arrangement at the boundaries which hold fixed and are not updated
during the iterations [Fig. 4b].

It is important to establish that our analytical model reproduces the
results obtained by numerical simulations. On a lattice the in-plane angles $%
\varphi _n$ deviate from the formula (\ref{logasol}) and obtain
modifications largest near the core spin. These angles satisfy a discrete
Laplace equation 
\begin{equation}
\sum\limits_\delta \sin (\varphi _n-\varphi _{n+\delta })=0  \label{Laplace}
\end{equation}
issued from Eq.(\ref{System2}).

Similar to the analytical prediction, we find that the radial dependence of
the arrangement is mostly preserved, $\varphi _n$ is determined only by an
absolute distance $r$ measured from the core site ($N,N$) beginning with $%
r\sim 3$. Figure 4c compares the numerical calculations performed for
different scan directions with the analytical values given by Eq.(\ref
{logasol}). The points in the inset are fitted by $\varphi =\pi
/2+0.135\,\ln \left( \frac r{10.84}\right) $. The mean-field approximation (%
\ref{alphaMF}) would give $\alpha =2/\sqrt{17}\pi \approx 0.154$, where the
magnetic field is $g\mu h/(JS)=1$. One sees that the agreement is very good.

It makes sense to explore on a lattice the dependence $\alpha [\varphi (\vec{%
r}_0)]$. Instead of inclusion of local in-plane magnetic field the following
simple but effective scheme was applied to enforce a desired logarithmic
source position. Again a square lattice is used with a fixed spin at its
center. This core spin with a given $\varphi (\vec{r}_0)$ would be excluded
from updating in the numerical routine. In order to find a range of values
for the coefficient $\alpha $ we investigate systems of size $21\times 21$
and $41\times 41$. Figure 5a summarizes the results found from the relaxed
configurations $\varphi _{ij}$ with different starting deviations $\varphi (%
\vec{r}_0)$ by showing $\alpha $ as a function of $\varphi (\vec{r}_0)$. One
sees the dependence is almost linear. The numerical $\alpha $ values for $%
N=21,41$ are slightly different, i.e. $\alpha $ weakly depends on the
lattice size. The results indicate that relevant $\alpha $ values are small,
i.e. $\left| \alpha \right| <1$. More significantly, the analytical model
only contains two parameters, $\alpha $ and $R$, and a relaxed spin
configuration, obtained numerically, can only depend upon two independent
variables $\varphi (\vec{r}_0)$ and $N$.

We then used the numerical code to calculate the energy of the structure 
\begin{equation}
\frac E{JS^2}=\frac 12\sum\limits_{np}\cos \left( \varphi _n-\varphi
_p\right) +\sum\limits_n\cos \left( \varphi _n-\varphi _0\right) -E_0,
\label{enertot}
\end{equation}
where the first sum runs over all inner sites, the sum in the second term
goes over the boundary sites that belongs to the structure. The energy was
calculated relative to the ground-state energy $E_0$, an amount of $JS^2$
per exchange bond, for spins aligned with an uniform angle $\varphi _0$
within the $xy$ plane. The energy found as a function of $\alpha ^2$, taken
from the fitting, is linear right up to a maximal value $\alpha _{\max }$
corresponding to $\varphi (\vec{r}_0)=\pi \,$that well agrees with the
continuum model (Fig.5b).

\subsubsection{Pair of logarithmic sources}

It is interesting to confirm the results found above by analysis an assembly
of such kind of logarithmic sources. According to the continuum model, the
simplest formation is a configuration including two sources of strength $%
\alpha $ and $-\alpha $

\begin{equation}
\varphi =\frac \alpha 2\ln \frac{\left( x-x_1\right) ^2+\left( y-y_1\right)
^2}{R_1^2}-\frac \alpha 2\ln \frac{\left( x-x_2\right) ^2+\left(
y-y_2\right) ^2}{R_2^2}=\phi _0+\frac \alpha 2\ln \frac{\left( x-x_1\right)
^2+\left( y-y_1\right) ^2}{\left( x-x_2\right) ^2+\left( y-y_2\right) ^2},
\label{pairphi}
\end{equation}
where $\phi _0=\alpha \ln \left( R_2/R_1\right) $. The energy of the pair 
\begin{equation}
E=2\pi \alpha ^2JS^2\ln d  \label{pairene}
\end{equation}
is finite and have no dependence on system size.

We use a square system of size $2N+1$ with two sources placed symmetrically
near its center which has the coordinates ($N,N$). We found in the numerical
studies that only distance between the sources and their mutual orientation
on a lattice affect the values of energy and $\alpha $. We present here our
results obtained from simulations when the sources are placed at positions ($%
N-d/2,N$) and ($N+d/2,N$), where $d$ is a distance between them.
Alternatively, if the sources were placed at different positions away from
the system center a boundary energy that changes significantly with the pair
positions would result. To avoid this complication, it is much simpler to
fix the source positions at the system center.

We investigate how the strength of source $\alpha $ could depend on the pair
distance $d$ and on a difference between the core spin angles $\delta
\varphi =\varphi (r_{20})-\varphi (r_{10})$ and how the analytical
predictions for the energy ($E\sim \alpha ^2$, $E\sim \ln d$) are modified
on a lattice. The calculation required a larger system than for the previous
case in order to produce stable configuration in a finite scale. We checked
the finite size effects by simulating $21\times 21$ to $1001\times 1001$
square lattices. In addition, in the calculations presented here we control
a restoring of homogeneous arrangement at system edges. The analytical
solution (\ref{pairphi}) involves two independent parameters, i.e. the
strength $\alpha $ and the distance $d$. In numerical simulations a number
of independent quantities involved in the calculation is the same. The first
parameter is determined by the difference $\delta \varphi $ and the second
one is given explicitly via the source positions.

It is of importance to determine an optimal size of the system to avoid
impractically long simulations for large enough lattice. In the Table I some
of the data, answering the purpose, is collected. The binding energies, the
strength of source $\alpha $ and a relaxed background arrangement $\varphi
_0 $ are summarized there. As a starting configuration we take two spins at
the distance $d=4$ turned almost oppositely to another magnetic moments. We
run then iteration procedure to obtain equilibrium configuration. Fitting to
the known solution (\ref{pairphi}) allows the determination of the needful
quantities. One see that the background arrangement $\varphi _0$ does not
approach the exact value $\pi /2 $ when decreasing the size of the lattice ($%
L\leq 101$). Size effects are noticed for the strength of source $\alpha $
which becomes greater for small lattices. Increasing the $L$ from $201$ till 
$301$ has an effect neither on $\alpha $ nor on $\varphi _0$. As a rule of
thumb, we hold the finite-size effects are negligible when $L\geq 50d$.

\begin{table}[tbp]
\caption{Data of numerical simulations for pair of logarithmic sources.}
\label{tableI}%
\begin{ruledtabular}
\begin{tabular}{cccc}
{Size} & {$E/JS^2$} & {$\alpha$} & {$\varphi _0$} \\
\hline \\
$21\times 21$ & $15.9945$ & $0.203$ & $1.562$ \\
$51\times 51$ & $15.9984$ & $0.127$ & $1.569$ \\
$31\times 31$ & $15.9973$ & $0.160$ & $1.567$ \\
$101\times 101$ & $15.9988$ & $0.108$ & $1.5703$ \\
$201\times 201$ & $15.9989$ & $0.1005$ & $1.5707$ \\
$301\times 301$ & $15.9989$ & $0.1005$ & $1.5707$%
\end{tabular}
\end{ruledtabular}
\end{table}

To compare the analytical expression (\ref{pairphi}) with a target in-plane
arrangement (Fig. 6) we choose a scan along a path in the $(0,1)$ direction
of the lattice, beginning with the point of one of the source. Similar
results could be obtained along other scan directions, but with a worse
consistence with the analytical predictions, since the region, where the
continuum model is expected to be valid, is not isotropic. A fit of data
points according to Eqs.(\ref{pairphi}, \ref{pairene}) provides estimates
for the parameter $\alpha $ and the energy $E$.

The dependence $\alpha $ on $\delta \varphi $ was examined for several $d$
values. We found that the observed dependence is distinguished for small ($%
d=2\div 4$) and large ($d\geq 6$) pair distances.

For sufficiently small distances ($d\leq 4$) the pair presents an unit
formation and cannot be treated adequately by the continuum model. A typical
result for the energy obtained for a square lattice of size $L=141$ is shown
in Fig. 7. The parameter $\alpha $ found as a function of $\delta \varphi $
is shown in Fig. 7a. One sees that the dependence exhibits a clear
periodical behavior. Due to the fact, the dependence $E(\alpha ^2)$
calculated with the same data and plotted in Fig. 7b turns out to be
nonlinear and many-valued.

By contrast, for the large distances ($d>10$) the agreement between the
analytical model and the numerical simulations ($L=701$) is good enough. In
Fig. 8a we show $\alpha $ as a function of $\delta \varphi $. The dependence
is almost linear and this behavior is similarly observed for one logarithmic
source. The energy of the pair as a function of $\alpha ^2\,$ also supports
the agreement, $E$ is directly proportional to $\alpha ^2$ until $\delta
\varphi \simeq \pi $ (Fig. 8b).

\subsubsection{ Nodal solutions of Heisenberg model.}

The patterns which we have so far studied have been confined in the $xy$%
-plane. A particularly interesting and complex case is when the structure
takes an out-of-plane form. Here we report numerical simulations of nodal
states. The lattice is taken as in Fig.3, however, in the initial
configuration the pinned spin in the center has an out-of-plane component.
The numerical procedure recurs those used for one logarithmic source and
generates both a set $\left( \cos \varphi _{ij},\sin \varphi _{ij}\right) $
and $\left( \cos \theta _{ij},\sin \theta _{ij}\right) $. This allows us to
extract from these simulations parameters involved in analytical expressions
such as those derived in Sec.I.

Given a set of angles found numerically, we should summarize the data by
fitting it to the model (\ref{targets}) that depends on the adjustable
parameters $\varphi _0$, $c$, $\alpha $ and $R$, and the fit supplies the
appropriate coefficients. To carry out the comparison we follow two steps.
From Eqs.(\ref{targets}) we obtain 
\begin{equation}
\sqrt{c^2-1}\cos \varphi _0\cos \varphi _{ij}+\sqrt{c^2-1}\sin \varphi
_0\sin \varphi _{ij}=\tan \theta _{ij}.  \label{mod1}
\end{equation}
Fitting Eq.(\ref{mod1}) to the resulting spin structure by the least-square
method, we obtain $c$ and $\varphi _0$. The rest parameters $\alpha $ and $R$
are then adjusted to data for $\cos \theta \,$. We estimate 
\[
\varphi _{ij}=\tan ^{-1}\left[ c\tan \left( \alpha \ln \left[ \frac{r_{ij}}R%
\right] \right) +\varphi _0\right] 
\]
with found $\alpha ,$ $R$, $c$ and $\varphi _0$ and compare $\varphi _{ij}$
with numerical data points (Fig.9). One sees, the agreement is rather well.

A set of numerical simulations was performed using different initial
conditions for boundary $\vec{S}_b=\left( \sin \theta _b\cos \varphi _b,\sin
\theta _b\sin \varphi _b,\cos \theta _b\right) $ spins and the central
pinned spin $\vec{S}_c=\left( \sin \theta _c\cos \varphi _c,\sin \theta
_c\sin \varphi _c,\cos \theta _c\right) $. We also used the expressions (\ref
{targets}) as fitting formulae. These fits give us values for the strength $%
\alpha $ and the energy $E$ listed in Table II. We found a significant
signature: the $\alpha $ value is determined only by the angle $\triangle
_{bc}=\cos ^{-1}\left( \vec{S}_b\vec{S}_c\right) $ for any direction $\vec{S}%
_c$. To check this assertion we repeat calculations of Sec.(III.B.1) for the
pure in-plane case with the starting deviations $\varphi (\vec{r}%
_0)=\triangle _{bc}$. A fit of data points according to Eqs.(\ref{logasol},%
\ref{enertot}) gives the estimates for strength of source $\alpha $ and the
energy $E$ (last two columns in the Table II) which are close to the
out-of-plane values. In addition, these simulations confirm the analytical
results for the energy: $E\sim \alpha ^2$ and $E$ does not depend on $c$.

\begin{table}[tbp]
\caption{Data of numerical simulations for nodal states.}
\label{tableIII}%
\begin{ruledtabular}
\begin{tabular}{cccccccccc}
$\delta \varphi =\varphi _c-\varphi _b$ & $\delta \theta =\theta _c-%
\frac \pi 2$ & $\triangle _{bc}$ & $E/JS^2$ & $c$ & $\varphi _0$ & $R$ & $\alpha $ & $\alpha$ (plane) & $E$ (plane) \\
\hline \\
$1.5$ & $0.5$ & $1.509$ & $1.513$ & $1.140$ & $-\frac \pi 2$ & $21.54$ & $%
-0.317$ & $-0.318$ & $1.513$ \\
$1.5$ & $-0.5$ & $1.509$ & $1.513$ & $1.140$ & $-\frac \pi 2$ & $21.54$ & $%
-0.317$ & $-0.318$ & $1.513$ \\
$0.5$ & $1.0$ & $1.077$ & $0.744$ & $3.399$ & $\frac \pi 2$ & $29.62$ & $%
-0.227$ & $-0.228$ & $0.714$ \\
$2.5$ & $1.5$ & $1.627$ & $1.758$ & $23.58$ & $\frac \pi 2$ & $21.69$ & $%
-0.340$ & $-0.340$ & $1.757$ \\
$0$ & $1.509$ & $1.509$ & $1.513$ & $\infty $ & $-$ & $21.54$ & $-0.317$ & $%
-0.318$ & $1.513$%
\end{tabular}
\end{ruledtabular}
\end{table}

From these features we could conclude that a pinned spin in the center is a
reason of appearance of logarithmic solutions both in XY and Heisenberg
models.

\subsection{Spiral vortex}

In the present section we continue with simulations of a spiral vortex, and
begin with XY case. It is meaningful to investigate whether the approach
described earlier keeps its validity for simulations of an ideal vortex. To
perform numerical simulations we consider a square lattice of even size $%
2N\times 2N$ shown in Fig.10a and place the vortex core in the center of the
dual lattice. Then, we start with an initial configuration with $q\neq 0$
and impose free boundary condition. After 5000 iterations we reach a relaxed
configuration shown in Fig.10b. On a lattice, the in-plane vortex angles $%
\varphi _{ij}$ lose the perfect circular symmetry of this formula, and
obtain modifications largest near the vortex core with the coordinates $(%
\tilde{i}_0,\tilde{j}_0)$. To check a validity of the solutions 
\begin{equation}
\cos \varphi _{ij}=\frac{i-\tilde{i}_0}{\sqrt{(i-\tilde{i}_0)^2+(j-\tilde{j}%
_0)^2}},\sin \varphi _{ij}=\frac{j-\tilde{j}_0}{\sqrt{(i-\tilde{i}_0)^2+(j-%
\tilde{j}_0)^2}}  \label{latcoord}
\end{equation}
we control a fulfillment of a discrete lattice nonlinear Laplace equation 
\[
\sum\limits_n\sin \left( \varphi _p-\varphi _n\right) =0, 
\]
where $n$ runs over the nearest neighbors of the $p$-th site, with an
accuracy of order $10^{-2}\,$ in the center and $10^{-5}$ at the outskirts.

As a next step, we perform numerical simulations for a spiral vortex
determined by the most common expression for the harmonic function 
\[
a(x,y)=q\arctan \frac{y-\tilde{y}_o}{x-\tilde{x}_o}+\alpha \ln \frac{\sqrt{%
(x-x_0)^2+(y-y_0)^2}}R. 
\]
Here we briefly review how this procedure has been organized. Assuming the
vortex is placed at some position centered in a plaquette and a logarithmic
source position coincides with one of the lattice sites the parameters $%
\tilde{x}_o,\tilde{y}_o$ $(x_0,y_0)$ are chosen in the dual (direct)
lattice. Firstly, we consider an initial vortex configuration, where one of
the spins nearest to the vortex is considered as a site of logarithmic
source. The in-plane angle of this core site is not changed by the iteration
procedure, and a fixed boundary condition holds. Farther from the core the
relaxed spin angles must well be described by the continuum formula $\varphi
(x,y)=q\phi +\alpha \ln \left( r/R\right) $, then we expect the spin
configuration 
\[
\varphi _{ij}=q\arctan \frac{j-N+1/2}{i-N+1/2}+\alpha \ln \frac{\sqrt{%
(i-x_0)^2+(j-y_0)^2}}R 
\]
would be seen to develop.

We investigate system of size $101\times 101$. The starting deviation $%
\varphi (\vec{r}_0)$ of the core spin in the uniform background results in
the relaxed configuration with $\alpha $ and the energy $E$ (Fig.11a). Most
importantly, we observe a logarithmic dependence of the in-plane angles $%
\varphi _{ij}$ for scans (used further in the fitting to determine $\alpha $
values) along paths in definite directions (Fig.11b) while scans along
another directions shows the opposite feature (Fig.11c). This procedure has
been repeated at several different positions of the fixed spin ($x_0$, $y_0$%
) and the results are shown in Table III, where in the last column we show
the $\alpha $ values derived from the calculated energy $E$.

\begin{table}[tbp]
\caption{Data of numerical simulations for spiral vortex.}
\label{tableV}%
\begin{ruledtabular}
\begin{tabular}{cccc}
($x_0$, $y_0$) & $(E-E_{vortex})/JS^2$ & $\alpha $ & $\left| \alpha \right|$ (energy) \\
\hline \\
$(N,N)$ & $0.114$ & $-0.084$ & $0.0828$ \\
$(N+5,N)$ & $0.554$ & $-0.179$ & $0.183$ \\
$(N+10,N)$ & $0.562$ & $-0.182$ & $0.184$ \\
$(N+15,N)$ & $0.568$ & $-0.189$ & $0.185$%
\end{tabular}
\end{ruledtabular}
\end{table}
This is done by using the expression $E=E_{\log }+E_{vort},$where 
\[
E_{\log }=\pi JS^2{\alpha }^2\ln \frac La,\,\,\,\,E_{vort}=\pi JS^2q^2\ln 
\frac La, 
\]
for the spiral vortex energy in the continuum approximation which is
suggested to be valid. The value for $E_{vort}/JS^2=16.58$ is taken from the
simulations of the ideal vortex. With this general expression, where $%
E_{vort}/E_{\log }=q^2/\alpha ^2$, it is easy to find $\alpha $ which will
be given by $\alpha (\mbox{energy})=\sqrt{\left( E-E_{vort}\right) /E_{vort}}
$. The following features are evident from Table III.

1) The farther the fixed spin from a vortex center the closer an additional
energy $E-E_{vort}$ (second column) to value found for a fixed spin in the
uniform background. The region of continuum theoretical description
decreases that is in agreement with the simulations for two opposite
logarithmic sources already reported in Sec.III.B.2.

2)\ The strength $\alpha $ depends both on $\varphi (\vec{r}_0)$ value and a
position of the fixed spin. The closer latter to a vortex center the less $%
\alpha $ at fixed $\varphi (\vec{r}_0)$ value. Far from the vortex source
the parameter $\alpha $ becomes exactly equal to value found for one
logarithmic source with the same lattice size and starting deviation $%
\varphi (\vec{r}_0)$.

3) The prediction of continuum theory $E\sim \left( \alpha ^2+q^2\right) $
is full reproduced by simulations for any logarithmic source and vortex
positions. The lowest energy value is obtained for the smallest distance
between the vortex center and the fixed spin.

\subsection{Space spiral vortex}

We begin with the case $\alpha =0$. To obtain this kind of spin structures
we solved the equations (\ref{sinpn}, \ref{cospn}, \ref{costp}) by first
setting the angles to their continuum values (\ref{outvort}) and then
iteratively setting each spin components to point along the direction of the
effective field due to its neighbors. However, an attempt to obtain the
space vortex by this way on the full square or disk fails. For these systems
the iteration procedure either converges to a Skyrmion structure (free
boundary conditions) or does not converge at all when the boundary spins
held fixed. The reason is the iteration procedure relaxes to a minimal
energy state. Since the feature of a starting configuration employed in
numerical calculations is a non-zero angular momentum, this configuration
may evolve either into the Skyrmion-like or into the space vortex-like
structures. However, the former has a gain in energy in comparison with the
last one. To avoid this difficulty we consider a simple way in which the
Skyrmion structure loses the advantage. Noting that an essential
contribution to the vortex energy comes from the spins within a small-radius
core we takes the computional region in the form of ring with the inner
radius $R_1$ and the external radius $R_2$. This results in small
differences of the discrete solution from the continuum result (\ref{outvort}%
).

We found the energy for the different parameters $c$ as shown in Table IV
for the ring of size $R_1=50.5$ and $R_2=105.5$ with free boundary
conditions.

\begin{table}[tbp]
\caption{Data of numerical simulations for space vortex.}
\label{tableVI}%
\begin{ruledtabular}
\begin{tabular}{cc}
$c$ & $E$ \\
\hline \\
1.0 & 2.107 \\
1.5 & 2.107 \\
3.5 & 2.107
\end{tabular}
\end{ruledtabular}
\end{table}

In full agreement with the continuum theory these energies have no
dependence on $c$ values. In Table V\ we listed the space vortex energy as a
function of the inner radius $R_1$ at fixed $R_2=100.5$. We see that
agreement between the numerical and the continuum theory result $E=\pi
JS^2\ln \frac{R_2}{R_1}$ is nice for the whole range of radius $10.5\leq
R_1\leq 50.5$. We may conclude that in the system without a finite-radius
core the space vortex will have the lowest energy among solutions with a
non-zero angular momentum. This assertion is supported by direct analytical
consideration \cite{Bostrem}.

\begin{table}[tbp]
\caption{Space vortex energy as a function of the inner radius $R_1$.}
\label{tableVII}%
\begin{ruledtabular}
\begin{tabular}{ccc}
$R_1$ & $E$ (lattice) & $E = \pi JS^2\ln \frac{R_2}{R_1}$ \\
\hline \\
10.5 & 7.05 & 7.10 \\
20.5 & 4.93 & 5.00 \\
30.5 & 3.68 & 3.75 \\
40.5 & 2.80 & 2.86 \\
50.5 & 2.11 & 2.16
\end{tabular}
\end{ruledtabular}
\end{table}
Now we turn to the space spiral vortex with $\alpha \neq 0$. To obtain this
configuration spins belonging to both boundaries of the ring are held fixed.
The constants $\phi _0(R_1)$ and $\phi _0(R_2)$ should be taken different
(''twisted'' boundary conditions). Figure 12 presents the results of the
modeling for $R_1=50.5$, $R_2=105.5$ and $\phi _0(R_2)=\phi _0(R_1)+\pi /4$
resulting in a spiral vortex structure with the energy $E=2.37JS^2$.

Recently, the statics and dynamics of flat circular magnetic nanostructures
with an in-plane magnetic vortex configuration has been investigated within
the framework of the Landau-Lifshitz-Gilbert equation putting particular
emphasis on the polarization of the vortex center and on the in-plane
vorticity \cite{Killinger}. Studying fast switching process induced by
out-of-plane field pulses, the authors was no longer dealing with a vortex
state, but rather with a spiral (see figure 12 in \cite{Killinger}). They
found also that in nanorings with an inner radius $R_1$ and an outer radius $%
R_2$ the stability of the vortex state is enhanced, and concerning the
switching of the vorticity, the nanorings have similar properties as
circular ones, i.e. with $R_1=0$.

In summary, we have studied a new class of spiral vortex-type solutions in a
2D Heisenberg ferromagnet and performed numerical simulations for various
spiral vortex configurations using fixed twisted boundary conditions and
pinned core spins (''superinstanton'' boundary conditions). These
simulations show a reasonable agreement with the continuum-approximation
results. Based on the investigation we may identify among the nonlinear
excitations the modes with circular nodes (''nodal'' solutions \cite
{Bostrem1}) and modes with azimuthal nodes of magnetization $M_z$ (space
spiral vortex) that resembles the classification of magnetostatic modes
excited by a magnetic-field pulses and observed recently in micron-sized
ferromagnetic disks. Incidentally, only axially symmetric magnetostatic
modes appeared if the tipping pulse is uniform over the disk and all
geometries are perfectly axially symmetric. Symmetry breaking modes,
instead, required, e.g., a non-uniform tipping pulse having a sizable
gradient in the plane of the vortex or a deviation of the sample from a
perfect cylindrical shape \cite{PRL2005}. However, the frequency of
nonaxially symmetric magnetostatic modes has a negative dispersion, i.e. it
decreases with a growth of a number of azimuthal nodal lines. Unlike this,
for the nonlinear excitations, which are of exchange origin, a number of
these lines coincides with a number of spiral arms (or with the vortex
topological charge $q$) and increases the energy. These stationary nonlinear
modes must be taken into account for yielding a better understanding, e.g.,
the fast magnetic switching properties in magnetic memory materials.

\acknowledgments
We would like to thanks B.A. Ivanov and M.V. Sadovskii for the useful
discussions. This work was partly supported by the grant NREC-005 of US CRDF
(Civilian Research \& Development Foundation)and by the grant RFBR(N
03-01-00100).

\begin{figure}[tbp]
\caption{Spiral vortices with $\alpha =2.0\,$ [$q=1$ (a), $q=2$ (b)] and
with $\alpha =0.3\,$ [$q=1$ (c), $q=2$ (d)]. Density images of the amplitude
and the phase of the magnetization are given for the last set. The map (e)
has one node curve dividing the bright-dark regions of equal amplitude but
opposite phase. The map (f) consists of four regions, oscillating in pairs
in phase. Note a similarity of topological small-$\alpha $ spiral vortices
with dynamical magnetostatic nonaxially symmetric modes (see Fig. 1 in Ref.
[23]) and Figs. 3-4 in Ref. [21].}
\label{solutions}
\end{figure}
\begin{figure}[tbp]
\caption{Core spin (white arrow) points along an axis determined jointly by
the exchange field $zJS$ of the nearest neighbors and the local filed $h$. }
\label{solutions}
\end{figure}
\begin{figure}[tbp]
\caption{Coordinates ($i,j$) for square lattice. The core spin is denoted by
white-black circle. Solid circles indicate the inner sites involved into
iteration procedure. Open circles are the boundary sites.}
\label{solutions}
\end{figure}
\begin{figure}[tbp]
\caption{Starting configuration used to get a logarithmic source (a) and
equilibrium configuration obtained in the iteration procedure (b). Core spin
is placed at position ($10,10$). The final equilibrium state reached after $%
50000$ iterations with an accuracy $10^{-15}$. A comparison between the
numerical simulation (points) and the analytical model (lines). To exclude
the lattice artifacts due to fixed boundary conditions we use the points
with $1\leq \ln r\leq 2$ for the fitting shown in the inset (c). }
\label{solutions}
\end{figure}
\begin{figure}[tbp]
\caption{The $\alpha $ value as a function of $\varphi (r_0)$ (a); the
energy $E$ as a function of $\alpha ^2$ (b). Note that the maximal $\alpha $
values found for different lattice sizes, $\alpha _{\max }=0.650$ ($N=21$)
and $\alpha _{\max }=0.615$ ($N=41$), are close to the estimation $0.636$ of
the mean-field theory.}
\label{solutions}
\end{figure}
\begin{figure}[tbp]
\caption{Starting configuration used to get a pair of logarithmic sources
(a) and relaxed configuration obtained in the iteration procedure with an
accuracy $\sim 10^{-7}$ (b). In the starting configuration the in-plane
angles of the selected sites are set equal to $\pi /2+\delta \varphi /2$ and 
$\pi /2-\delta \varphi /2$ and they are hold fixed during the iteration
scheme [Eqs.(\ref{xy1},\ref{xy2})]. The spins at the edges of the system
have been included into the iteration procedure too and they have no
constraint from outside, that is a free boundary condition holds. For all
another spins, the in-plane angles are supposed to be $\pi /2$. }
\label{solutions}
\end{figure}
\begin{figure}[tbp]
\caption{Parameter $\alpha $ as a function of $\delta \varphi $ (a) and
energy $E$ as a function of $\alpha ^2$ (b) for a small-distance pair ($d=4$%
). The pair presents an unit formation and cannot be treated adequately by
the continuum model.}
\label{solutions}
\end{figure}
\begin{figure}[tbp]
\caption{Dependencies $\alpha (\delta \varphi )$ (a) and $E(\alpha ^2)$ (b)
for a large-distance pair ($d=10$). The agreement between the analytical
model and the numerical simulation is good enough.}
\label{solutions}
\end{figure}
\begin{figure}[tbp]
\caption{Comparison of $\cos \varphi $ obtained from numerical simulation
(black points) to the analytical expression $\cos \varphi =\cos \left\{
\arctan \left[ 1.140\tan \left( -0.307\ln r+2.544\right) \right] -\pi
/2\right\} $ given by the continuum theory (solid line). Core spin angles
are $\varphi (r_0)=1.5$ and $\theta (r_0)=0.5$. }
\label{solutions}
\end{figure}
\begin{figure}[tbp]
\caption{Starting configuration for a simulation of vortex structures on a
lattice of size ($2N$, $2N$) (a). The center with coordinates ($N-1/2$, $%
N-1/2$) is denoted by black circle. Relaxed configuration obtained after
5000 iterations (b). The estimation of energy $E/JS^2$ for system of size $%
100\times 100$ by using numerical codes $\varphi _{ij}$ [Eq.(\ref{enertot})]
is 16.58 vs 17.72 given by the continuum theory.}
\label{solutions}
\end{figure}
\begin{figure}[tbp]
\caption{Relaxed spiral-vortex configuration with the energy $E/JS^2=16.6994$%
. Pinned ($N,N$)-spin is denoted by the dotted circle (a). Dependencies of
the in-plane angles $\varphi _{ij}$ along the $(-1,-1)$ (b) and $(1,-1)$ (c)
directions with logarithmic and non-logarithmic behavior, respectively. }
\label{solutions}
\end{figure}
\begin{figure}[tbp]
\caption{Space spiral vortex (c=2): arrangement in the plane $xy$ (a) and
the $\cos \theta \,$ profile(b).}
\label{solutions}
\end{figure}

\end{document}